# Nucleocytoplasmic transport: a thermodynamic mechanism

Ronen Benjamine Kopito and Michael Elbaum*
Department of Materials and Interfaces
Weizmann Institute of Science
Rehovot 76100, Israel

tel 972 8 9343537
fax 972 8 9344138

*  author for correspondence
e-mail michael.elbaum@weizmann.ac.il

**<u>Abstract</u>**

The nuclear pore supports molecular communication between cytoplasm and nucleus in eukaryotic cells. Selective transport of proteins is mediated by soluble receptors, whose regulation by the small GTPase Ran leads to cargo accumulation in, or depletion from the nucleus, i.e., nuclear import or nuclear export. We consider the operation of this transport system by a combined analytical and experimental approach. Provocative predictions of a simple model were tested using cell-free nuclei reconstituted in *Xenopus* egg extract, a system well suited to quantitative studies. We found that accumulation capacity is limited, so that introduction of one import cargo leads to egress of another. Clearly, the pore *per se* does not determine transport directionality. Moreover, different cargo reach a similar ratio of nuclear to cytoplasmic concentration in steady-state. The model shows that this ratio should in fact be independent of the receptor-cargo affinity, though kinetics may be strongly influenced. Numerical conservation of the system components highlights a conflict between the observations and the popular concept of transport cycles. We suggest that chemical partitioning provides a framework to understand the capacity to generate concentration gradients by equilibration of the receptor-cargo intermediary.

**Introduction**

Enclosure of the genome within a nucleus is a defining feature of all eukaryotes. The spatial separation of the chromatin from the cytoplasm demands a system for regulated molecular communication. The portal for this traffic is the nuclear pore, a large protein channel that traverses both the outer and inner bilayer membranes that comprise the nuclear envelope (NE) (Schwartz TU, 2005; Alber F et al, 2007). The nuclear pore is typically present in multiple copies, a few hundred in yeast to several thousand in mammalian tissue culture cells. In amphibian oocytes their number may reach into the millions, while a recent electron tomographic study of the smallest known eukaryote, *O. taurus*, found only a few nuclear pores (Henderson GP et al, 2007). Three-dimensional reconstruction techniques have been applied to *Xenopus* oocytes, amoeba, and yeast, with a strong structural similarity in all species examined to date (Akey CW et al, 1993; Stoffler D et al, 1999; Beck M et al, 2007; Hinshaw JE et al, 1992; Yang Q et al, 1998). In addition, a nearly complete proteomic picture is available for yeast and human cells (Rout MP et al, 2000; Cronshaw JM et al, 2002). Nucleoporins of closely related species show very high sequence homology, while homology between nucleoporins of distant species may be as low as 20%. This sequence divergence has been used to trace evolutionary development of eukaryotes (Mans BJ et al, 2004; Bapteste E et al, 2005; Devos D et al, 2004; Denning DP et al, 2007) and may have reached a maximal level consistent with structural preservation. Most intriguingly, yeast carrying quite drastic temperature-sensitive modifications of the nuclear pores still maintained cell viability (Strawn LA et al, 2004; Zeitler B et al, 2004). These observations show that the basic operation of the nucleocytoplasmic transport system should be maintained across a great diversity of implementations.

As a biological channel, the nuclear pore is unique for the broad spectrum of macromolecular traffic it supports. Messenger RNA, ribosomal subunits, and transfer RNA must move from their sites of synthesis in the nucleus to the cytoplasm where proteins are translated. Conversely, proteins present in the nucleus must have entered it from the cytoplasm. Two modes of transport are distinguished in the literature. Small molecules, including water, ions, metabolites, and some proteins may pass the pore diffusively. This has traditionally been called "passive" transport. Conversely, the traffic of larger substrates is highly selective, and depends on interaction with soluble, intermediary receptors that usher them through the pore as a complex (Macara IG, 2001; Görlich D et al, 1999). These receptors typically recognize peptide signals on the cargo. Thus nuclear localization signals (NLS) govern nuclear import via interaction with "importin" family receptors, and nuclear export signals (NES) encode nuclear export through interaction with related "exportin" receptors. As suggested by the nomenclature, transport events are often considered to be irreversible; once delivered, a transport substrate should remain in the target compartment unless and until it is specifically returned by a receptor of the opposite sense. More than 20 such nucleocytoplasmic transport receptors (NTRs), also known as karyopherins, have been identified to date in humans (Pemberton LF et al, 2005). Cognate signals have been identified for a few of these, in particular the classical NLS types for importin $\alpha/\beta$ (Kalderon D et al, 1984; Robbins J et al, 1991), M9 and other signals for transportin (Pollard VW et al, 1996; Lee BJ et al, 2006), and NES for CRM1 (Fornerod M et al, 1997; Fukuda M et al, 1997; Ossareh-Nazari B et al, 1997; Stade K et al, 1997; Wen W et al, 1995). Because of the ability of the nuclear pore and associated receptors to move the substrate into the nucleus (or cytoplasm) against a gradient, and because of the dependence on nucleotide triphosphate hydrolysis, the underlying mechanism has been called "active" nucleocytoplasmic transport.

The interaction of NLS or NES cargo proteins with their cognate receptors is regulated by a small GTPase, Ran. Receptor-mediated transport is driven by a spatial gradient of RanGTP, which is high in the nucleus and low in the cytoplasm (Kalab P et al, 2002). This gradient is established by the localization of the guanosine exchange factor RanGEF, RCC1, in the nucleus, and the GTPase activating protein RanGAP on the cytoplasmic face of the nuclear pore itself. Transport receptors of different types bind RanGTP competitively or cooperatively with their cargo; the former lead to nuclear accumulation, the importins, and the latter to depletion, the exportins. These interactions are normally described in terms of transport "cycles". For nuclear "import", a receptor-cargo complex forms in the cytoplasm but is abrogated by RanGTP in the nucleus, leaving the cargo effectively stranded there. The receptor-RanGTP pair then returns to the cytoplasm, where the GTP is hydrolyzed and the receptor freed for another round of import. Conversely, for nuclear "export", a trimeric complex forms in the nucleus, passes the nuclear pore, and breaks up upon arrival in the cytoplasm when the GTP is hydrolyzed. The export receptor re-enters the nucleus empty of cargo. As both these receptor types essentially export RanGTP, a distinct receptor is required to maintain the nuclear level of Ran. NTF2 binds to RanGDP and releases it upon GTP exchange (Ribbeck K et al, 1998; Smith A et al, 1998; Steggerda SM et al, 2000). In the transport cycles picture, the RanGTP gradient is ultimately responsible for directional transport through the pore, and GTP hydrolysis is tightly coupled to translocation.

An intriguing addition to the picture above is that the major import receptor, importin $\beta1$, normally operates as a heterodimer with an adapter molecule, importin $\alpha$. Isoforms of $\alpha$ may then recognize different NLS, increasing the versatility of the $\beta1$ receptor (Köhler M et al, 1999). Additional roles for $\alpha$ have been proposed in regulating cargo transport (Riddick G et al, 2007). Importantly, $\alpha$ is auto-inhibited for cargo binding to conventional NLS in the absence of $\beta$ (Fanara P et al, 2000; Harreman MT et al, 2003; Fradin C et al, 2003; Catimel B et al, 2001), so that prior assembly of the heterodimer is required for binding. On the other hand, there is evidence that $\alpha$ crosses the nuclear pore without $\beta$ (Miyamoto Y et al, 2002), and may act independently as a transport receptor for certain substrates (Kotera I et al, 2005; Hübner S et al, 1999). In the nucleus, only importin β interacts directly with RanGTP. GTP binding releases the α-NLS complex; α then binds a different receptor, CAS, together with RanGTP. In this respect, CAS is an "exporter" for α. Thus in the "cycles" picture, two RanGTP molecules are normally required for the import of a single cargo by importin β1. Certain substrates bind directly to β1 without an adapter (Lam MH et al, 1999; Lee SJ et al, 2003; Cingolani G et al, 2002). Thus energy dissipation as GTP hydrolysis may be coupled one-to-one or two-to-one per translocation event.

In addition to regulation by Ran, the nuclear pore might possess some structural asymmetry that could contribute to the apparent directionality of transport through it. Several observations of retrograde transport lie at odds with such a suggestion, however (Schmidt-Zachmann MS et al, 1993; Nachury MV et al, 1999; Becskei A et al, 2003). Computer simulation shows that net accumulation may be achieved without a specific assumption of vectorial transport (Görlich D et al, 2003). We have recently published quantitative experimental data testing the machinery of nucleocytoplasmic transport in cell-free nuclei reconstituted from *Xenopus* egg extract (Kopito RB et al, 2007). This model system supports a transport analogy to classical enzymology, where cargo is converted from a cytoplasmic to a nuclear form (Stein WD, 1989). Our work justified a thermodynamic view of transport as a process of molecular partitioning based on equilibration of the receptor-cargo intermediate. We found that nuclear accumulation follows simple first-order kinetics, with a saturating level in the nucleus that depends on the concentration of cargo in the cytoplasm. The saturation curve represented a stable coexistence in the cargo distribution, in that the same conditions were reached by accumulation or by dilution of the cargo in the cytoplasm. At steady-state a balanced, fast flux of cargo was maintained in both directions across the nuclear pores, mediated by the "import" receptor. In contrast to any presumption of irreversible delivery, the pore machinery allowed the receptor-cargo complex to exchange freely between the nucleus and cytoplasm.

In this work, we explore the kinetic or thermodynamic operation of the nuclear pore by a combined approach of modeling and experiment. We first translate the Ran-regulation model into the language of chemical thermodynamics. The derivation makes a number of challenging predictions, which we then test experimentally. Specifically, we find a strong coupling between the accumulation of distinct import cargoes, which is consistent with a maximal capacity for the system as a whole. We show further that different cargoes addressing the same receptor reach the same nuclear to cytoplasmic concentration ratios in steady state. This indicates that the fluorescent probe indeed displays the state of the endogenous, non-fluorescent cargoes present as well in the cell extract. We then reexamine the thermodynamic model and the consequences of conservation in the numbers of transport factors. While NTRs and Ran move between the cytoplasmic and nuclear compartments, they are not created, nor destroyed, as a part of the cargo transport process. We find that RanGTP egress cannot rely exclusively on the import receptor, and that the common feature of Ran regulation leads to a coupling between cargo accumulation even where different receptors are involved. Together, the model and experiments offer a view of nucleocytoplasmic transport as an integrated cellular system.

**Results**

Basic model

In modeling, we consider the nuclear pore as a selective barrier. Its permeability for each substrate $S$ is denoted $p_S$. The concentration of RanGTP is $[R]$, and RanGDP $[R']$; the NTF2 receptor is $[N]$. $[C]$ denotes the concentration of an NLS-bearing cargo and $[T]$ the cognate nucleoctyoplasmic transport receptor (e.g., importin α/β dimer), or NTR. $[E]$ denote exportins, which carry cargo cooperatively with RanGTP. For simplicity we consider that $[E]$ collectively includes CAS. Superscripts $C$ and $N$ denote concentrations in the cytoplasm or in the nucleus, respectively. Subscripts $fr$ represent free factors in equilibrium with their potential complex partners. Table I lists the available interactions in the nuclear and cytoplasmic environments.

We first summarize our earlier phenomenological model (Kopito RB et al, 2007). NTRs and their cargoes interact via: $[T] + [C] \Leftrightarrow [TC]$. Since cytoplasmic Ran is in the GDP form ($[R]^C = 0$), $[TC]^C$ can reach a simple binding equilibrium in relation to $[C]^C$ with affinity $K$:

(1)     $$[TC]^C = \frac{[T_{tot}]^C \dfrac{[C]^C}{K}}{1 + \dfrac{[C]^C}{K}} \quad .$$

At the initial stage of accumulation ($[C]^N_{t=0} = 0$), the influx of $[TC]^C$ takes place into an environment that is empty of both $[C]^N$ and $[TC]^N$. The exchange reaction between receptor-bound cargo and RanGTP is expressed as:

(2)     $$[TC]^N + [R_{fr}]^N \Leftrightarrow [TR]^N + [C]^N \quad .$$

Even as $[C]^N$ begins to rise, $[TC]^N$ remains close to zero since the receptors have higher affinity to $[R]^N$. Nuclear cargo may bind to fixed sites, e.g. on chromatin, in which case it is lost from the reach of the transport system. Alternatively, the cargo protein may remain soluble in the nucleoplasm, perhaps after saturating its binding sites if the concentration suffices. Accumulation will progress until the nuclear cargo competes with RanGTP for binding the transport receptor. Detailed balance occurs when the higher affinity of RanGTP for the receptor is compensated by the higher concentration of the soluble nuclear cargo. Steady state is obtained when the concentrations of the NTR-cargo complex inside and outside of the nucleus are balanced: $[TC]^N_{SS} = [TC]^C$.

Endogenous transport cargo

When considering assays based on fluorescent probes introduced to cell extracts, the new cargo enters an environment that is crowded with endogenous NLS-cargoes. These may already have reached a steady-state distribution. A similar situation occurs in live cells where signaling or synthesis of a new nuclear protein occurs on the background of prior homeostasis, including both specific binding and non-specific competition. Transport assays that follow a single fluorescent tracer must take into account this background of endogenous factors, as NTRs may interact with a variety of different cellular cargoes. We can consider these in the framework of multiple substrates that compete for binding to a single enzyme (Dixon M et al, 1964; Haldane JBS, 1965). Specifically, multiple NLS cargoes, as well as RanGTP, bind to the NTR, which lowers the free energetic barrier for their passage through the nuclear pore. We may categorize the cargoes into one fluorescent tracer, $[C_x]$, with affinity to the receptor $K_x$, and the remaining endogenous cargoes indexed as $[C_e]$, each with receptor affinity $K_e$. The association of a receptor with a given cargo, in quasi-equilibrium, is given by the product of their concentrations divided by the mutual affinity, normalized by the affinity-weighted sum of all other components interacting with the same receptor. Local equilibration of the NTR-cargo complexes in the cytoplasm and the nucleus thus yields:

$$(3) \quad [TC_x]^C = \frac{[T_{tot}]^C \frac{[C_x]^C}{K_x}}{1 + \sum_{\{e\}} \frac{[C_e]^C}{K_e} + \frac{[C_x]^C}{K_x}} \qquad (4) \quad [TC]^N = \frac{[T_{tot}]^N \frac{[C_x]^N}{K_x}}{1 + \frac{[R_{fr}]^N}{K_R} + \sum_{\{e\}} \frac{[C_e]^N}{K_e} + \frac{[C_x]^N}{K_x}}$$

By equating equations (3) and (4), we obtain that the nuclear to cytoplasmic ratio at steady-state is equal for any cargo species:

$$(5) \quad \frac{[C_x]_{SS}^N}{[C_x]^C} = \frac{[C_e]_{SS}^N}{[C_e]^C} \ .$$

The index $x$ can in fact represent any cargo, where the set of all the others is considered endogenous. The relation then holds for any and all cargo species. The individual affinities for the receptor, $K$, do not appear in the ratio as the denominators present a sum over all cargo, endogenous and probe alike. The receptor affinity of the probe cargo, $K_x$, may affect the kinetics of its accumulation, as well as the value of the final steady-state ratio, but that value will be common to all cargoes.

If endogenous cargoes outnumber NTRs, so that most receptors are occupied, i.e., $\sum_{\{e\}} [TC_e]^C \approx [T_{tot}]^C$ , then an increment in the cargo concentration upon adding the probe will hardly affect the fraction of bound receptors, which is anyhow close to unity. However, each receptor-cargo species will experience some readjustment according to the relative affinities:

$$(6) \qquad \frac{[TC_x]^C}{[TC_e]^C} = \frac{[C_x]^C K_e}{[C_e]^C K_x} \ .$$

Whatever $[C_x]$ is added to the cytoplasm, it will cause a decrease in $[TC_e]^C$, and $[TC_e]^N$ will have to decrease in order to reestablish the new steady-state conditions. As a result, a chemical gradient in the endogenous cargo appears, in a direction opposite to that of the probe. While the probe accumulates in the nucleus, the thermodynamic model predicts that a compensatory efflux of receptors with endogenous cargoes should occur into the cytoplasm.

Experimental transport assays

In order to test these predictions we performed a set of experiments using two differently-labeled protein tracers: GFP-nucleoplasmin (GFP-NP), and BSA labeled chemically with synthetic SV40 large T-antigen NLS (BSA-NLS), and dyed with tetramethyl rhodamine. Nuclear accumulation assays were performed using cell-free nuclei reconstituted in *Xenopus* egg extract (Lohka MJ et al, 1983; Newmeyer DD et al, 1986). Molecular concentrations were calibrated by fluorescence correlation spectroscopy and probed dynamically by confocal fluorescence intensity. First, GFP-NP was introduced at some initial concentration and allowed to reach steady-state over a period of at least 90 min. BSA-NLS was then introduced to the cytosol at varying concentrations, from 230 nM to 3.6 μM, and the nuclear concentrations of both substrates were followed over time. The cytosolic concentrations remained constant due to the negligible volume fraction of the nuclei. In all cases the GFP-NP exited as BSA-NLS entered. Fig 1A shows two representative curves with GFP-NP and BSA-NLS concentrations as marked. Note that addition of one substrate causes an egress of the other, and that the two substrates reach similar nuclear to cytosolic ratios in steady state, as predicted. In Fig 1B the asymptotic steady-state $[C]^N/[C]^C$ ratios are shown for both substrates, where the cytosolic GFP-NP concentration was 600 nM. The two substrates reach approximately equal ratios for each cytosolic level of cytosolic BSA-NLS. For higher BSA-NLS concentrations, the common, steady-state nuclear to cytosolic ratio decreases. This suggests that the system of nuclear accumulation has a finite capacity that can be distributed among cargo substrates. When challenged with a high concentration in one specific cargo, the nuclear concentrations of all others adjust to accommodate.

In order to emphasize further the common nuclear to cytoplasmic ratio, we performed a different sort of comparative kinetic assay. First, GFP-NP and BSA-NLS were introduced to an import assay together, and their $[C]^N/[C]^C$ ratio was followed over time. Fig 2A shows that the two curves track each other accurately. The fact that the normalized curves behave similarly suggests that $p_{TC}$ is similar for the GFP-NP and the BSA-NLS substrates. Second, the same concentrations were employed but in the mode of Fig 1A. The titration order was reversed; BSA-NLS was added initially and allowed to reach steady-state, after which GFP-NP was added. As GFP-NP accumulated in the nucleus the BSA-NLS substrate moved outward, shown in Fig 2B. Notably, the two substrates reached the same steady-state value seen in Fig 2A. These data further confirm the thermodynamic operation of the nucleocytoplasmic transport system; the endpoint is independent of the kinetic path taken to reach it.

If cargo added to the cytoplasm can reach the nucleus only after binding a transport receptor, then the kinetics with which receptors redistribute to cargo in the cytoplasm should have a determining influence on the kinetics of nuclear accumulation. In the background of endogenous cargoes, one would expect relatively fast accumulation if the probe affinity is stronger than the endogenous, and vice versa. Eqn. 6 provides a framework to predict this partitioning. The initial rate of accumulation is the product of the permeability $p_{TC}$ and the concentration $[TC_x]^C$, with the important caveat that establishment of the equilibrium distribution of receptors and cargoes is presumed. A more complete development appears in Supplementary Information, along with a comparison to data of a recent paper in which nuclear accumulation kinetics were measured in yeast cells for a range NLS-receptor affinities (Hodel AE et al, 2006). The prediction and the measurements show a striking qualitative agreement.

<u>Conservation of transport factors</u>

In order to better understand the system operation we return to the thermodynamic model. In the mathematical expression we can specifically require that transport factors are conserved on the relevant time scale; i.e., they are exchanged between the nucleus and cytoplasm, but otherwise they are neither created nor destroyed. We begin with Ran. It is small enough to pass the nuclear pore autonomously, but its flux is enhanced by association with receptors, [$N$] for RanGDP, and [$T$] or [$E$] for RanGTP. We presume that the activities of RanGEF and RanGAP suffice to maintain nuclear Ran as RanGTP and cytoplasmic Ran as RanGDP. In the cytoplasm, [$T$] and [$E$] do not interact with RanGDP, while $[N]^C + [R']^C \Leftrightarrow [NR']^C$. In the nucleus, [$N$] does not interact with RanGTP. Hypothetically, in the absence of cargo the receptors would equilibrate individually with nuclear RanGTP: $[T]^N + [R]^N \Leftrightarrow [TR]^N$ (or $[E]^N + [R]^N \Leftrightarrow [ER]^N$). The receptor-RanGTP complexes pass back to the cytoplasm with a permeability $p_{TR}$ ($p_{ER}$) where they dissociate upon hydrolysis of the GTP and conversion of [$R$] to [$R'$]. As [$TR$] ([$ER$]) complexes are absent from the cytoplasm their efflux rate is simply proportional to their nuclear concentration. Similarly, the influx of [$NR'$]$^C$ is proportional to its concentration in the cytoplasm. We can thus write the total influx of RanGDP and efflux of RanGTP as:

(7)     $J_{R'} = p_{R'}[R'_{fr}]^C + p_{NR'}[NR']^C$ ,

(8)     $J_R = p_R[R_{fr}]^N + p_{TR}[TR]^N + p_{ER}[ER]^N$ .

The Ran system that drives nucleocytoplasmic transport can reach steady-state only if these fluxes are equal.

$J_{R'}$ and $J_R$ are plotted schematically in Fig 3A along horizontal axes of opposing orientation for $[R]$ and $[R']$, considering first a closed cell with finite volume. The total Ran concentration is conserved in their sum anywhere along this axis. The shapes of the curves reflect possible binding equilibria and proportions of autonomous and receptor-mediated transport. (The lower dashed lines show autonomous transport only, where flux is simply proportional to concentration in each compartment.) The intersection of the curves defines a working point for the Ran system where influx and efflux are in detailed balance. In the special case of a large cytoplasmic reservoir, Fig. 3B, $[N]^C$ and $[R']^C$ are unaffected by nuclear transport, and therefore $J_{R'}$ is also constant, albeit with a specific value that depends on details. This defines a single working point to which $J_R$ must accommodate by adjustment of $[R]^N$, $[TR]^N$, and $[ER]^N$, which allows for a family of hypothetical curves intersecting the working point.

In addition to Ran, the flux of NTRs passing the nuclear pore must come to detailed balance in order to support a steady state behavior. In particular, the flux balance of $[T]$ is expressed as:

(9)     $p_T[T_{fr}]^C + p_{TC}[TC]^C = p_T[T_{fr}]^N + p_{TR}[TR]^N + p_{TC}[TC]^N$ .

With $[TC]^C = [TC]^N$ at steady-state,

(10)     $ss:$     $p_{TR}[TR]^N = p_T([T_{fr}]^C - [T_{fr}]^N) = p_T \Delta T_{fr}$ .

When cargoes are more abundant than receptors, and the concentrations of both exceed the equilibrium affinity between them, the vast majority of receptors will be bound. Therefore $[T_{fr}]$ will be very small in both compartments, and the difference $\Delta T_{fr}$ still smaller. Inevitably, in steady state the contribution to the RanGTP flux by "import" receptors $[T]$ is close to zero; $p_{TR}[TR]^N \approx 0$ and either the pore permeability or the nuclear concentration of $[TR]^N$ must be vanishingly small. The remaining terms in Eqns. 7 and 8 for the Ran flux balance represent autonomous transport and egress in complex with exportins, $[E]$. In that case, we have

(11)     $ss:$     $p_{R'}[R'_{fr}]^C + p_{NR'}[NR']^C \approx p_{ER}[ER]^N + p_R[R_{fr}]^N$ .

At steady state, NTF2-mediated import of RanGDP is balanced primarily by so-called export receptors carrying RanGTP.

Coupling of independent pathways

We next consider that the single Ran system serves multiple NTRs associated with distinct families of NLS. This suggests the possibility of coupling between different pathways, similar to the coupling between accumulation of different substrates of the same receptor. Aside from importin α/β1, the best characterized is the transportin pathway (Pollard VW et al, 1996; Lee SJ et al, 2003). Therefore we performed the competition assay of Fig 1 using BSA-NLS and a GFP fusion to the transportin substrate hnRNP A1 (GFP-A1). GFP-A1 was allowed to reach steady-state, and then BSA-NLS was added. Results appear in Fig 4, where we see that indeed the nuclear concentration of transportin substrate decreases as the importin substrate accumulates. Nuclear RanGTP must partition between the various NTRs ( $[T_1]$ , $[T_2]$ , etc), and addition of a new cargo leads to a repartitioning equivalent to a move between the different curves shown schematically in Fig 3B. In steady state the net flux of Ran must still be balanced with $J_{R'} = J_R$ . Thus, indirectly, addition of cargo directed to one NTR affects the ability of nuclear Ran to compete with $[TC]^N$ for all receptors. The increased load thus causes a general reduction in nuclear accumulation of cargo.

## Discussion

The cellular subsystem for regulated nucleocytoplasmic transport may be viewed according to two fundamentally different paradigms. The first is essentially kinetic. The nuclear pore is considered in this light as an active molecular transporter, and specific interactions taking place at the pore should determine its transport properties. Nuclear "import" and "export" arise as natural concepts if the pore is capable of vectorial delivery. As the nomenclature suggests, an importin carries its cargo in and an exportin carries it out. This paradigm inspires the mechanistic dissection of interactions between receptors and the pore, including models based on affinity gradients (Pyhtila B et al, 2003; Ben-Efraim I et al, 2001) or sliding peristalsis (Melcák I et al, 2007). An alternate viewpoint is more thermodynamic. The nuclear pore is then one part of a system that operates as a molecular pump, or machine, for accumulation of nuclear protein cargoes. The pore provides selectivity in this picture by restricting translocation to legitimate receptor complexes.

Directionality of translocation remains a matter of current debate in the literature. On one hand, continuous shuttling between the nucleus and cytoplasm is traditionally thought to require both NLS and NES sequences (Hodel AE et al, 2006; Riddick G et al, 2007). On the other hand, as discussed in the introduction there are recurring reports of retrograde transport, and we have shown that NLS cargo moves bidirectionally with its receptor. Steady-state represents a balanced flux, i.e., a continuous exchange (Kopito RB et al, 2007). We must emphasize that assays relying on fluorescence intensity alone cannot detect this exchange, which occurs even when the nuclear concentration is substantially higher than the cytoplasmic. In a completely independent context, photobleaching of the Bicoid morphogen in *Drosophila* embryo nuclei showed repeated, fast recovery consistent with bidirectional shuttling (Gregor T et al, 2007). Single-molecule studies provide the most direct view of interactions at the pore (Yang W et al, 2004; Kubitscheck U et al, 2005; Tokunaga M et al, 2008). Bidirectional transport has been recognized as "abortive events" in single-molecule observation (Yang W et al, 2006). The enigma is resolved in recognizing that "import" refers to net nuclear accumulation of a protein population, rather than targeted delivery of the individual molecules. Classical nuclear import thus strongly resembles what has been known as "nuclear shuttling" (Michael WM, 2000).

The present study was motivated by the thermodynamic paradigm. We first built a simple, first principles model of regulated, receptor-mediated transport. A realistic comparison to nucleocytoplasmic transport requires two explicit constraints. The first accounts for the background of endogenous, cellular cargoes that compete with the fluorescent probe in the transport system. The second is conservation of the system components. The latter is equivalent to the notion that nucleocytoplasmic transport does not depend on synthesis or degradation of factors other than GTP. The new model made a number of provocative predictions. We set out to test these experimentally, and found them well justified. In particular, the model suggested that the steady-state nuclear to cytoplasmic concentration ratio of an import substrate should be independent of its affinity to the transport receptor. Therefore different proteins addressing the same receptor should reach similar ratios. By extension, the ratio observed with a fluorescent probe should represent the corresponding ratio of endogenous cargoes as well. Note that this applies only to the soluble fraction of any given protein; binding to immobile structures sequesters the substrate from the transport system. (Again, fluorescence intensity measures the sum of soluble and immobilized substrate; alone, it may provide a misleading measure of the transport process, whose evaluation requires a dynamic technique such as fluorescence correlation spectroscopy or fluorescence recovery after photobleaching.) Moreover, the total capacity for nuclear accumulation of all substrates should be finite, suggesting that import of one cargo should cause an egress of all others. In fact we found that this coupling between substrates occurs even for cargoes that address different transport receptors. This was quite unanticipated, but reflects the fact that a single Ran system regulates the entire importin β family.

Thermodynamically, it is reasonable that a given rate of GTP to GDP conversion can maintain only a finite cargo concentration in the nucleoplasm. GTP to GDP conversion is expressed in our model by the outward flux of RanGTP, $J_R$. Conservation of Ran demands that $J_R$ is equal to the inward flux of RanGDP, $J_{R'}$. Demanding conservation of NTRs in addition to Ran led to a paradox: we could not account simultaneously for the steady-state behavior observed, together with recycling of RanGTP to the cytoplasm by the NTR itself. Another exit pathway for RanGTP was required. While this could be supplied by the overall complement of exportin receptors, it is worth recalling that the major transport receptor importin α/β normally operates as a heterodimer. The β unit is often considered the biochemical receptor per se, while the α molecule acts as the adapter to the NLS. In fact, α normally binds to NLS only when already bound to β (Catimel B et al, 2001; Kobe B, 1999; Harreman MT et al, 2003; Fanara P et al, 2000; Fradin C et al, 2005), so that the receptor for classical NLS cargo is properly the heterodimer. After their dissociation, α may return to the cytoplasm with CAS and RanGTP, and β should also return to the cytoplasm with RanGTP. This suggests that an essential role for α is in sustaining the balanced flux of Ran. Substrates that bind to importin β directly see essentially a distinct receptor. Direct binding might hinder assembly of the heterodimer. This is probably the case for the synthetic importin β binding domain of importin α, a popular substrate in transport assays.

The thermodynamic paradigm might be regarded as an extension of "transport cycles", where the coupling of hydrolysis to cargo delivery is satisfied only statistically. Indeed the two models converge in the particular limit of infinite RanGTP. By contrast, all indications from our results suggest that RanGTP is limiting. In spite of its higher affinity to the importin family receptors, it must compete against all soluble cargoes that accumulate in the nucleus. A detailed simulation showed that elevated RanGEF (RCC1) may lead to sequestering of Ran in the nucleus, and that this may even inhibit NLS-cargo import (Riddick G et al, 2005). Another simulation predicted the possibility that cargo competition to RanGTP could lead to bidirectional exchange, though the consequences were not fully explored (Görlich D et al, 2003). The question of Ran recycling also distinguishes the present paradigm from that of transport cycles. We found that mediation of Ran recycling by import receptors alone violates their conservation; it must be complemented by exportins or CAS. Interestingly, it was observed that recycling of importin β can occur independently of Ran (Kose S et al, 1999); this corresponds to the [$T_{fr}$] term in our flux balance equations The coupling that we have observed between behavior of different import substrates is explained very naturally by the saturation of their common regulator in a system whose behavior depends primarily on chemical partitioning.

We are not able to reconcile the present findings with a purely kinetic view of the nuclear pore. Addition of one "import" substrate after another could hardly be expected to "export" the first. The convergence of nuclear to cytoplasmic ratios for different substrates, one accumulating and the other diluting, is quite unthinkable if "import" is governed locally at the pore. It is also hard to imagine the observed coupling between importin β and transportin substrates in a paradigm that focuses solely on their arrival from one side of the pore. We conclude that the model, the data, and the agreement between them strongly support a thermodynamic view of nucleocytoplasmic transport as a cellular system based on chemical partitioning. Conditions in the cytoplasmic and nuclear volumes surrounding the nuclear envelope determine the kinetics and direction of net substrate translocation, and the constraints of conservation couple between cellular import pathways in subtle ways. The essential role of the nuclear pore is as a selective filter for receptors and their complexes. The basic physical principles that provide this local selectivity are a subject of intense, independent investigations (Rout MP et al, 2000; Ribbeck K et al, 2002; Frey S et al, 2007; Macara IG, 2001; Peters R, 2005; Zilman A et al, 2007; Bickel T et al, 2002; Kustanovich T et al, 2004; Kapon R et al, 2008; Naim B et al, 2007). To some degree it can be reconstituted synthetically (Caspi Y et al, 2008; Jovanovic-Talisman T et al, 2008; Frey S et al, 2007). In the physiological situation, cargo translocation certainly involves an intricate set of nucleoporin-receptor interactions at the pore, some of which may involve also RanGTP (Lyman SK et al, 2002; Nakielny S et al, 1999; Shah S et al, 1998). Recent single-molecule measurements by microinjection to living cells (Dange T et al, 2008) showed that receptor-mediated translocation of the pore is rapid, with a narrow distribution of dwell times comparable to *in vitro* measurements. This finding is consistent with the present data in indicating that local intermediates at the pore appear not to play a direction-determining or rate limiting role in the integrated transport system behavior, i.e., in kinetics or steady-state accumulation of NLS-cargo.

Experiments were performed in a particular, perhaps peculiar environment with an essentially infinite cytoplasm. *In vitro* studies with permeabilized cells are similar in this respect (Adam SA et al, 1990). The large cytoplasm provides convenient conditions in which to study transport, and lessons about the role of the pore itself can be extended to other cellular implementations. The basic considerations presented here are equally valid for cells with a confined cytoplasm. The major difference is expressed in the working point of the Ran balance as shown in Fig 3. For a large cytoplasmic reservoir, the concentrations of all molecules and molecular complexes there are independent of their nuclear accumulation. In cells, by contrast, nuclear accumulation depletes the cytoplasm. Furthermore, the protein stock is in a continuous state of expression and degradation even in homeostasis. To the extent that this turnover is independent of the transport mechanism, however, i.e., that Ran or receptor degradation is not required in order to transport cargo, the constraint of conservation remains in force and the lessons of the model remain the same. Experiments by other groups in live cells have found rather similar, approximately first-order kinetics, typically with modest nuclear to cytoplasmic ratios (Rihs HP et al, 1991; Hu W et al, 2005; Timney BL et al, 2006; Riddick G et al, 2005). Another recent work addressed the saturating ratio issue specifically, comparing direct and α-mediated binding to importin β (Riddick G et al, 2007). While they found distinctly differing ratios in steady-state, the probe cargoes included an engineered NES so a direct comparison is not possible. One study in yeast provided a comprehensive scan of NLS affinities, and reported, without explanation, a complex relation between affinity and the nuclear to cytoplasmic ratio measured in an unsynchronized culture (Hodel AE et al, 2006). Our thermodynamic model provides an independent prediction for the dependence of accumulation kinetics on receptor-cargo affinity. The kinetics reflect the redistribution of endogenous cargoes to available receptors in the cytoplasm when a new substrate is added.

Accumulation kinetics will be fast if the affinity is higher than a weighted average of the endogenous cargo, and slow if the affinity is weaker. The model's prediction is qualitatively very similar to the prior observations in yeast.

Ran has been assigned the role of transport directionality regulator. On the other hand, we have shown that net nuclear accumulation does not imply directional transport of individual molecular substrates. What advantage, then, does Ran regulation provide over simple receptor-mediated transport without a competitive regulator? The latter process is also able in principle to transport substrates from one compartment to the other; upon interaction with the target site the receptor-cargo bond is lost. Ran regulation may provide fast kinetics for a wide range of substrates, as has been suggested (Stewart M, 2007). We suggest that a major feature of Ran regulation, rather than imposing a strict directionality through the nuclear pore, is in permitting nuclear accumulation of substrates in soluble form (Phair RD et al, 2000; Pederson T, 2000; Grünwald D et al, 2008; Costa M et al, 2006). This provides a means for scanning nuclear binding sites with weak affinities by transient interaction (Gorski SA et al, 2006). The target sites do not need to release the transport receptor. Soluble protein also provides a material bridge between the nucleoplasm and the cytoplasm, and a direct means for molecular communication between the two cellular compartments. This may have far-reaching consequences for signaling, cell fate, and development (Coppey M et al, 2007).

## Materials and Methods

Tetramethyl rhodamine labeled BSA-NLS was prepared as in (Salman H et al, 2001). GFP-nucleoplasmin and clarified crude *Xenopus* egg extract for nuclear reconstitution were prepared as in (Kopito RB et al, 2007). The measurement apparatus was a custom-built microscope with water-immersion optics (Zeiss C-Apochromat X40/1.2) incorporating differential interference contrast and epi-fluorescence imaging with fluorescence correlation spectroscopy using a pair of avalanche photodiodes (Perkin Elmer, Inc.). The instrument was modified for two color collection using Ar ion (488 nm) and green HeNe (543 nm) lasers coupled into a single optical fiber. Emission filters (Chroma, Inc.) were well separated spectrally (530/30 nm, and 610/75 nm). Fluorescent substrate concentrations were calibrated individually by fluorescence correlation spectroscopy. Cross-talk between the blue excitation and the red emission channel was eliminated by a linear transformation after calibration of the individual emission intensities for each substrate, and verified by testing of known mixtures.

Nuclei were assembled by mixing the egg extract with demembranated *Xenopus* sperm nuclei and an ATP regeneration system, and incubating for 90 minutes. Nuclei were visualized by differential interference contrast microscopy. Competitive import assays were performed by introducing the first fluorescent substrate to two separate assembly reactions. One was introduced to the microscope immediately in order to follow accumulation until steady-state was attained, typically 90 minutes. The second fluorescent substrate was then introduced to the second reaction, and transferred quickly to the microscope for kinetic measurement. Once a new steady-state was reached, the sample was scanned for nearby nuclei. The steady-state nuclear to cytoplasmic substrate ratio was then measured for at least 10 nuclei.

The model presented here is analytical in nature. It was analyzed using Matlab (Wolfram Research), and plots were prepared using Origin (OriginLab).

## Acknowledgments

The authors are grateful to A. Harel (Technion, Israel) for a gift of demembranated sperm nuclei. This work was supported in part by grants from the Israel Science Foundation and the Minerva Research Foundation, by the Estate of Rose L. Ban, and by the Gerhardt M.J. Schmidt Center for Supramolecular Architecture. This research is made possible in part by the historic generosity of the Harold Perlman Family.

## References

Adam SA, RS Marr, and L Gerace (1990) Nuclear protein import in permeabilized mammalian cells requires soluble cytoplasmic factors. *J Cell Biol* 111, 807-16

Akey CW, and M Radermacher (1993) Architecture of the Xenopus nuclear pore complex revealed by three-dimensional cryo-electron microscopy. *J Cell Biol* 122, 1-19

Alber F et al. (2007) The molecular architecture of the nuclear pore complex. *Nature* 450, 695-701

Bapteste E, RL Charlebois, D MacLeod, and C Brochier (2005) The two tempos of nuclear pore complex evolution: highly adapting proteins in an ancient frozen structure. *Genome Biol* 6, R85

Beck M, V Lucić, F Förster, W Baumeister, and O Medalia (2007) Snapshots of nuclear pore complexes in action captured by cryo-electron tomography. *Nature* 449, 611-5

Becskei A, and IW Mattaj (2003) The strategy for coupling the RanGTP gradient to nuclear protein export. *Proc Natl Acad Sci U S A* 100, 1717-22

Ben-Efraim I, and L Gerace (2001) Gradient of increasing affinity of importin beta for nucleoporins along the pathway of nuclear import. *J Cell Biol* 152, 411-7

Bickel T, and R Bruinsma (2002) The nuclear pore complex mystery and anomalous diffusion in reversible gels. *Biophys J* 83, 3079-87

Caspi Y, D Zbaida, H Cohen, and M Elbaum (2008) Synthetic Mimic of Selective Transport Through the Nuclear Pore Complex. *Nano Lett* 8, 3728-34

Catimel B, T Teh, MR Fontes, IG Jennings, DA Jans, GJ Howlett, EC Nice, and B Kobe (2001) Biophysical characterization of interactions involving importin-alpha during nuclear import. *J Biol Chem* 276, 34189-98

Cingolani G, J Bednenko, MT Gillespie, and L Gerace (2002) Molecular basis for the recognition of a nonclassical nuclear localization signal by importin beta. *Mol Cell* 10, 1345-53

Coppey M, AM Berezhkovskii, Y Kim, AN Boettiger, and SY Shvartsman (2007) Modeling the bicoid gradient: diffusion and reversible nuclear trapping of a stable protein. *Dev Biol* 312, 623-30

Costa M, M Marchi, F Cardarelli, A Roy, F Beltram, L Maffei, and GM Ratto (2006) Dynamic regulation of ERK2 nuclear translocation and mobility in living cells. *J Cell Sci* 119, 4952-63

Cronshaw JM, AN Krutchinsky, W Zhang, BT Chait, and MJ Matunis (2002) Proteomic analysis of the mammalian nuclear pore complex. *J Cell Biol* 158, 915-27

Dange T, D Grünwald, A Grünwald, R Peters, and U Kubitscheck (2008) Autonomy and robustness of translocation through the nuclear pore complex: a single-molecule study. *J Cell Biol* 183, 77-86

Denning DP, and MF Rexach (2007) Rapid evolution exposes the boundaries of domain structure and function in natively unfolded FG nucleoporins. *Mol Cell Proteomics* 6, 272-82

Devos D, S Dokudovskaya, F Alber, R Williams, BT Chait, A Sali, and MP Rout (2004) Components of coated vesicles and nuclear pore complexes share a common molecular architecture. *PLoS Biol* 2, e380

Dixon M, and EC Webb (1964) The Enzymes. 2nd ed. New York: Academic Press.

Fanara P, MR Hodel, AH Corbett, and AE Hodel (2000) Quantitative analysis of nuclear localization signal (NLS)-importin alpha interaction through fluorescence depolarization. Evidence for auto-inhibitory regulation of NLS binding. *J Biol Chem* 275, 21218-23

Fornerod M, M Ohno, M Yoshida, and IW Mattaj (1997) CRM1 is an export receptor for leucine-

rich nuclear export signals. *Cell* 90, 1051-60

Fradin C, A Abu-Arish, R Granek, and M Elbaum (2003) Fluorescence correlation spectroscopy close to a fluctuating membrane. *Biophys J* 84, 2005-20

Fradin C, D Zbaida, and M Elbaum (2005) Dissociation of nuclear import cargo complexes by the protein Ran: a fluorescence correlation spectroscopy study. *C R Biol* 328, 1073-82

Frey S, and D Görlich (2007) A saturated FG-repeat hydrogel can reproduce the permeability properties of nuclear pore complexes. Cell 130, 512-23

Fukuda M, S Asano, T Nakamura, M Adachi, M Yoshida, M Yanagida, and E Nishida (1997) CRM1 is responsible for intracellular transport mediated by the nuclear export signal. *Nature* 390, 308-11

Görlich D, and U Kutay (1999) Transport between the cell nucleus and the cytoplasm. *Annu Rev Cell Dev Biol* 15, 607-60

Görlich D, MJ Seewald, and K Ribbeck (2003) Characterization of Ran-driven cargo transport and the RanGTPase system by kinetic measurements and computer simulation. *EMBO J* 22, 1088-100

Gorski SA, M Dundr, and T Misteli (2006) The road much traveled: trafficking in the cell nucleus. *Curr Opin Cell Biol* 18, 284-90

Grünwald D, RM Martin, V Buschmann, DP Bazett-Jones, H Leonhardt, U Kubitscheck, and MC Cardoso (2008) Probing intranuclear environments at the single-molecule level. *Biophys J* 94, 2847-58

Gregor T, EF Wieschaus, AP McGregor, W Bialek, and DW Tank (2007) Stability and nuclear dynamics of the bicoid morphogen gradient. *Cell* 130, 141-52

Hübner S, HM Smith, W Hu, CK Chan, HP Rihs, BM Paschal, NV Raikhel, and DA Jans (1999) Plant importin alpha binds nuclear localization sequences with high affinity and can mediate nuclear import independent of importin beta. *J Biol Chem* 274, 22610-7

Haldane JBS (1965) Enzymes. MIT Press.

Harreman MT, MR Hodel, P Fanara, AE Hodel, and AH Corbett (2003) The auto-inhibitory function of importin alpha is essential in vivo. *J Biol Chem* 278, 5854-63

Henderson GP, L Gan, and GJ Jensen (2007) 3-D ultrastructure of O. tauri: electron cryotomography of an entire eukaryotic cell. *PLoS ONE* 2, e749

Hinshaw JE, BO Carragher, and RA Milligan (1992) Architecture and design of the nuclear pore complex. *Cell* 69, 1133-41

Hodel AE, MT Harreman, KF Pulliam, ME Harben, JS Holmes, MR Hodel, KM Berland, and AH Corbett (2006) Nuclear localization signal receptor affinity correlates with in vivo localization in Saccharomyces cerevisiae. *J Biol Chem* 281, 23545-56

Hu W, BE Kemp, and DA Jans (2005) Kinetic properties of nuclear transport conferred by the retinoblastoma (Rb) NLS. *J Cell Biochem* 95, 782-93

Jovanovic-Talisman T, J Tetenbaum-Novatt, AS McKenney, A Zilman, R Peters, MP Rout, and BT Chait (2008) Artificial nanopores that mimic the transport selectivity of the nuclear pore complex. *Nature*, 457, 1023-1027

Köhler M, C Speck, M Christiansen, FR Bischoff, S Prehn, H Haller, D Görlich, and E Hartmann (1999) Evidence for distinct substrate specificities of importin alpha family members in nuclear protein import. *Mol Cell Biol* 19, 7782-91

Kalab P, K Weis, and R Heald (2002) Visualization of a Ran-GTP gradient in interphase and mitotic Xenopus egg extracts. *Science* 295, 2452-6

Kalderon D, BL Roberts, WD Richardson, and AE Smith (1984) A short amino acid sequence able to specify nuclear location. *Cell* 39, 499-509

Kapon R, A Topchik, D Mukamel, and Z Reich (2008) A possible mechanism for self-coordination of bidirectional traffic across nuclear pores. *Phys Biol* 5, 36001

Kobe B (1999) Autoinhibition by an internal nuclear localization signal revealed by the crystal structure of mammalian importin alpha. *Nat Struct Biol* 6, 388-97

Kopito RB, and M Elbaum (2007) Reversibility in nucleocytoplasmic transport. *Proc Natl Acad Sci U S A* 104, 12743-8

Kose S, N Imamoto, T Tachibana, M Yoshida, and Y Yoneda (1999) beta-subunit of nuclear pore-targeting complex (importin-beta) can be exported from the nucleus in a Ran-independent manner. *J Biol Chem* 274, 3946-52

Kotera I, T Sekimoto, Y Miyamoto, T Saiwaki, E Nagoshi, H Sakagami, H Kondo, and Y Yoneda (2005) Importin alpha transports CaMKIV to the nucleus without utilizing importin beta. *EMBO J* 24, 942-51

Kubitscheck U, D Grünwald, A Hoekstra, D Rohleder, T Kues, JP Siebrasse, and R Peters (2005) Nuclear transport of single molecules: dwell times at the nuclear pore complex. *J Cell Biol* 168, 233-43

Kustanovich T, and Y Rabin (2004) Metastable network model of protein transport through nuclear pores. *Biophys J* 86, 2008-16

Lam MH, LJ Briggs, W Hu, TJ Martin, MT Gillespie, and DA Jans (1999) Importin beta recognizes parathyroid hormone-related protein with high affinity and mediates its nuclear import in the absence of importin alpha. *J Biol Chem* 274, 7391-8

Lee BJ, AE Cansizoglu, KE Süel, TH Louis, Z Zhang, and YM Chook (2006) Rules for nuclear localization sequence recognition by karyopherin beta 2. *Cell* 126, 543-58

Lee SJ et al. (2003) The structure of importin-beta bound to SREBP-2: nuclear import of a transcription factor. *Science* 302, 1571-5

Lohka MJ, and Y Masui (1983) Formation in vitro of sperm pronuclei and mitotic chromosomes induced by amphibian ooplasmic components. *Science* 220, 719-21

Lyman SK, T Guan, J Bednenko, H Wodrich, and L Gerace (2002) Influence of cargo size on Ran and energy requirements for nuclear protein import. *J Cell Biol* 159, 55-67

Macara IG (2001) Transport into and out of the nucleus. *Microbiol Mol Biol Rev* 65, 570-94

Mans BJ, V Anantharaman, L Aravind, and EV Koonin (2004) Comparative genomics, evolution and origins of the nuclear envelope and nuclear pore complex. *Cell Cycle* 3, 1612-37

Melcák I, A Hoelz, and G Blobel (2007) Structure of Nup58/45 suggests flexible nuclear pore diameter by intermolecular sliding. *Science* 315, 1729-32

Michael WM (2000) Nucleocytoplasmic shuttling signals: two for the price of one. *Trends Cell Biol* 10, 46-50

Miyamoto Y, M Hieda, MT Harreman, M Fukumoto, T Saiwaki, AE Hodel, AH Corbett, and Y Yoneda (2002) Importin alpha can migrate into the nucleus in an importin beta- and Ran-independent manner. *EMBO J* 21, 5833-42

Nachury MV, and K Weis (1999) The direction of transport through the nuclear pore can be inverted. *Proc Natl Acad Sci U S A* 96, 9622-7

Naim B, V Brumfeld, R Kapon, V Kiss, R Nevo, and Z Reich (2007) Passive and facilitated transport in nuclear pore complexes is largely uncoupled. *J Biol Chem* 282, 3881-8

Nakielny S, S Shaikh, B Burke, and G Dreyfuss (1999) Nup153 is an M9-containing mobile nucleoporin with a novel Ran-binding domain. *EMBO J* 18, 1982–1995

Newmeyer DD, DR Finlay, and DJ Forbes (1986) In vitro transport of a fluorescent nuclear protein and exclusion of non-nuclear proteins. *J Cell Biol* 103, 2091-102

Ossareh-Nazari B, F Bachelerie, and C Dargemont (1997) Evidence for a role of CRM1 in signal-mediated nuclear protein export. *Science* 278, 141-4

Pederson T (2000) Diffusional protein transport within the nucleus: a message in the medium . *Nat Cell Biol* 2, E73-E74

Pemberton LF, and BM Paschal (2005) Mechanisms of receptor-mediated nuclear import and nuclear export. *Traffic* 6, 187-98

Peters R (2005) Translocation through the nuclear pore complex: selectivity and speed by reduction-of-dimensionality. *Traffic* 6, 421-7

Phair RD, and T Misteli (2000) High mobility of proteins in the mammalian cell nucleus. *Nature* 404, 604-9

Pollard VW, WM Michael, S Nakielny, MC Siomi, F Wang, and G Dreyfuss (1996) A novel receptor-mediated nuclear protein import pathway. *Cell* 86, 985-94

Pyhtila B, and M Rexach (2003) A gradient of affinity for the karyopherin Kap95p along the yeast nuclear pore complex. *J Biol Chem* 278, 42699-709

Ribbeck K, G Lipowsky, HM Kent, M Stewart, and D Görlich (1998) NTF2 mediates nuclear import of Ran. *EMBO J* 17, 6587-98

Ribbeck K, and D Görlich (2002) The permeability barrier of nuclear pore complexes appears to operate via hydrophobic exclusion. *EMBO J* 21, 2664-71

Riddick G, and IG Macara (2007) The adapter importin-alpha provides flexible control of nuclear import at the expense of efficiency. *Mol Syst Biol* 3, 118

Riddick G, and IG Macara (2005) A systems analysis of importin-{alpha}-{beta} mediated nuclear protein import. *J Cell Biol* 168, 1027-38

Rihs HP, DA Jans, H Fan, and R Peters (1991) The rate of nuclear cytoplasmic protein transport is determined by the casein kinase II site flanking the nuclear localization sequence of the SV40 T-antigen. *EMBO J* 10, 633-9

Robbins J, SM Dilworth, RA Laskey, and C Dingwall (1991) Two interdependent basic domains in nucleoplasmin nuclear targeting sequence: identification of a class of bipartite nuclear targeting sequence. *Cell* 64, 615-23

Rout MP, JD Aitchison, A Suprapto, K Hjertaas, Y Zhao, and BT Chait (2000) The yeast nuclear pore complex: composition, architecture, and transport mechanism. *J Cell Biol* 148, 635-51

Salman H, D Zbaida, Y Rabin, D Chatenay, and M Elbaum (2001) Kinetics and mechanism of DNA uptake into the cell nucleus. *Proc Natl Acad Sci U S A* 98, 7247-52

Schmidt-Zachmann MS, C Dargemont, LC Kühn, and EA Nigg (1993) Nuclear export of proteins: the role of nuclear retention. *Cell* 74, 493-504

Schwartz TU (2005) Modularity within the architecture of the nuclear pore complex. *Curr Opin Struct Biol* 15, 221-6

Shah S, and DJ Forbes (1998) Separate nuclear import pathways converge on the nucleoporin Nup153 and can be dissected with dominant-negative inhibitors. *Curr Biol* 8, 1376-86

Smith A, A Brownawell, and IG Macara (1998) Nuclear import of Ran is mediated by the transport factor NTF2. *Curr Biol* 8, 1403-6

Stade K, CS Ford, C Guthrie, and K Weis (1997) Exportin 1 (Crm1p) is an essential nuclear export factor. *Cell* 90, 1041-50

Steggerda SM, BE Black, and BM Paschal (2000) Monoclonal antibodies to NTF2 inhibit nuclear protein import by preventing nuclear translocation of the GTPase Ran. *Mol Biol Cell* 11, 703-19

Stein WD (1989) Kinetics of transport: analyzing, testing, and characterizing models using kinetic approaches. *Methods Enzymol* 171, 23-62

Stewart M (2007) Molecular mechanism of the nuclear protein import cycle. *Nat Rev Mol Cell Biol* 8, 195-208

Stoffler D, B Fahrenkrog, and U Aebi (1999) The nuclear pore complex: from molecular architecture to functional dynamics. *Curr Opin Cell Biol* 11, 391-401

Strawn LA, T Shen, N Shulga, DS Goldfarb, and SR Wente (2004) Minimal nuclear pore complexes define FG repeat domains essential for transport. *Nat Cell Biol* 6, 197-206

Timney BL, J Tetenbaum-Novatt, DS Agate, R Williams, W Zhang, BT Chait, and MP Rout (2006) Simple kinetic relationships and nonspecific competition govern nuclear import rates in vivo. *J Cell Biol* 175, 579-93

Tokunaga M, N Imamoto, and K Sakata-Sogawa (2008) Highly inclined thin illumination enables clear single-molecule imaging in cells. *Nat Methods* 5, 159-61

Wen W, JL Meinkoth, RY Tsien, and SS Taylor (1995) Identification of a signal for rapid export of proteins from the nucleus. *Cell* 82, 463-73

Yang Q, MP Rout, and CW Akey (1998) Three-dimensional architecture of the isolated yeast nuclear pore complex: functional and evolutionary implications. *Mol Cell* 1, 223-34

Yang W, J Gelles, and SM Musser (2004) Imaging of single-molecule translocation through nuclear pore complexes. *Proc Natl Acad Sci U S A* 101, 12887-92

Yang W, and SM Musser (2006) Nuclear import time and transport efficiency depend on importin beta concentration. *J Cell Biol* 174, 951-61

Zeitler B, and K Weis (2004) The FG-repeat asymmetry of the nuclear pore complex is dispensable for bulk nucleocytoplasmic transport in vivo. *J Cell Biol* 167, 583-90

Zilman A, S Di Talia, BT Chait, MP Rout, and MO Magnasco (2007) Efficiency, selectivity, and robustness of nucleocytoplasmic transport. *PLoS Comput Biol* 3, e125

**Figure & Table Legends**

Figure 1.     Accumulation of one nuclear "import" cargo drives the efflux of a second. In steady-state both cargoes reach similar nuclear to cytoplasmic concentration ratios. (A) GFP-nucleoplasmin (GFP-NP) was allowed to accumulate until steady-state, and then a TRITC-labeled bovine serum albumin, chemically labeled with SV40 large T-antigen nuclear localization signals (BSA-NLS), was introduced. As BSA-NLS accumulated (red curves), the nuclear concentration of GFP-NP (green curves) was depleted. For the assay, GFP-NP represents one component of the endogenous protein population ( $[C_e]$ ), while the BSA-NLS is introduced as the probe ( $[C_x]$ ). Concentrations as marked. The fits show first-order kinetics. (B) Steady-state concentrations were measured for a constant initial concentration of GFP-NP (600 nM) and progressively increasing concentrations of BSA-NLS as marked along the horizontal axis. Both substrates reach similar values of $[C]^N/[C]^C$ for each BSA-NLS concentration. The fit is to the form

$$\frac{[C]^N}{[C]^C} = \frac{A}{B + [C]^C}$$

. The inset shows the same data plotted as $[C]^N$ vs. $[C]^C$ , and the fit in the Michaelis-Menten form following (Kopito RB et al, 2007).

Figure 2.     Nuclear accumulation reflects a thermodynamic target. (A) GFP-NP and BSA-NLS substrates were introduced to the import assay together. The kinetics of their accumulation are very similar. The red curve shows a first-order fit to the BSA-NLS. Concentrations as marked. (B) BSA-NLS was first allowed to accumulate to steady-state, and then GFP-NLS was introduced. Note that the order is opposite that of Figure 1. Again, the initial substrate exits the nucleus when challenged by the second cargo, consistent with a finite total capacity for nuclear accumulation. Additionally, both substrates reach similar ratios $[C]^N/[C]^C$ when steady-state is restored, and this ratio is equal to that shown in panel A. This demonstrates that the steady-state is a thermodynamic endpoint that may be reached by different kinetic paths.

Figure 3.     Conservation of Ran. The inward flux of Ran as RanGDP, $J_{R'}$ , depends on the level of cytoplasmic RanGDP and availability of NTF2 receptors. The outward flux of Ran as RanGTP, $J_R$ , depends on the nuclear concentration of free RanGTP, as well as "import" receptors $[T]^N$ and "export" receptors $[E]^N$ . Dissipation of chemical energy takes the form of GTP hydrolysis on Ran, i.e., the conversion of RanGTP to RanGDP. The rate of dissipation is thus proportional to $J_R$ . (A) For a cell of closed volume, the total number of Ran is conserved. A binding equilibrium between $[R']^C$ and $[N]^C$ leads to a flux $J_{R'} = p_{R'}[R']^C + p_{NR'}[NR']^C$ ,

shown in red with hypothetical parameters for permeabilities and affinities. A similar curve is shown in green for $J_R = p_R [R]^N + p_{TR}[TR]^N + p_{ER}[ER]^N$. Note that the nuclear RanGTP concentration $[R]^N$ is plotted ascending to the left. The intersection of the curves represents the working point for the transport system where total Ran is conserved. For comparison, the dotted lines represent autonomous flux of Ran, without facilitation by receptors. The total quantity of Ran is conserved and equal for any vertical line (dashed, in blue). For the example, concentrations are shown for a nuclear volume one fourth of the total cell. (B) In the case of a large cytoplasmic reservoir, or syncytium, $[R']^C$ and $[N]^C$ remain unaffected by transport since the nuclear volume is negligible. Thus $J_{R'}$ and $J_R$ must remain balanced. Nuclear RanGTP may nonetheless repartition among the free form $[R]^N$ and complexes with receptors, represented by the set of green curves. Addition of cargo may affect this balance.

Figure 4.    Competition between cargoes of distinct receptors. A GFP hybrid of hnRNP A1 protein, a transportin substrate, was first allowed to accumulate to steady-state, and then TRITC BSA-NLS was introduced. The concentration of the first decreased while the second accumulated. The two pathways are coupled through the nuclear RanGTP, which must alter its distribution among the available receptors in order to balance the RanGDP influx.

Table 1.    Protein interactions involved in the nucleocytoplasmic transport of protein cargoes bearing nuclear localization signals (NLS). The small GTPase Ran exists in GDP ($R'$) and GTP ($R$) states, governed by the auxiliary factors RanGAP in the cytoplasm and RanGEF (RCC1) in the nucleus. Transport receptors facilitate the passage of bound cargo through the nuclear pore: generically importins ($T$) and exportins ($E$). Importins bind their NLS-cargo ($C$) competitively with RanGTP, exportins cooperatively with RanGTP. Both essentially deplete the nucleus of RanGTP, while neither $T$ nor $E$ interacts with RanGDP. Ran may exchange autonomously, with conversion to GTP or GDP form on arrival; RanGDP transport is facilitated by NTF2 ($N$). Arrows for the nuclear pore represent the direction of flux: one-sided when the substrate exists only in the *cis* compartment, double-sided in the case of *cis-trans* exchange. Net flux of each substrate is simply proportional to its concentration difference with a set of permeabilities $p$. The nuclear pore is not considered to impose any preferential directionality. It is worth noting that the major receptor importin α/β1 binds NLS-cargo as a heterodimer. While this study focuses on nuclear accumulation, or "import", the cytoplasmic return of importin α alone is facilitated at least in part by the dedicated exportin CAS, in complex with RanGTP (not shown).

**Figures**

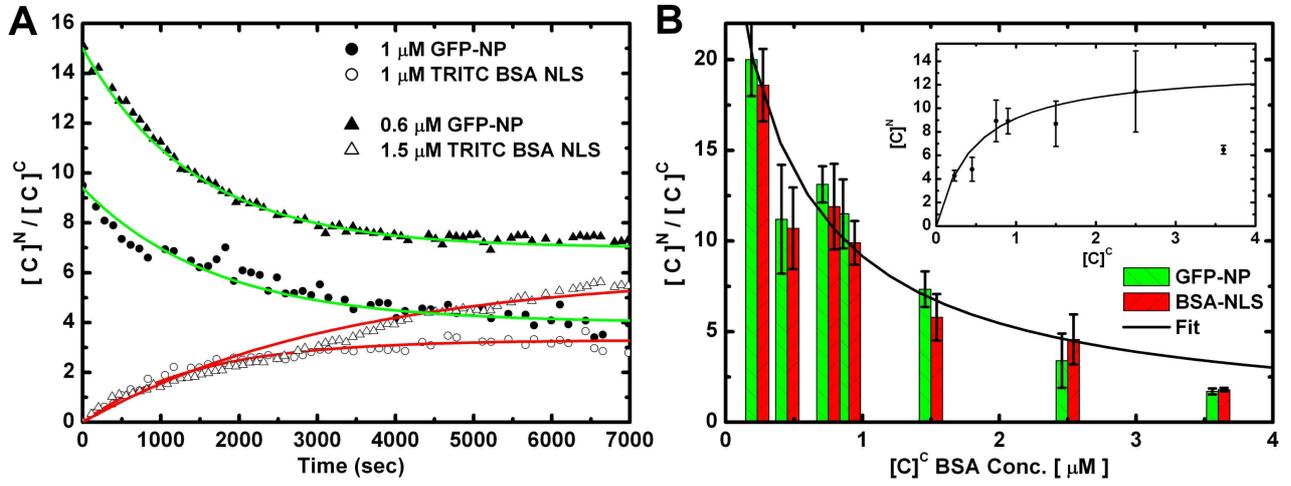

Figure 1

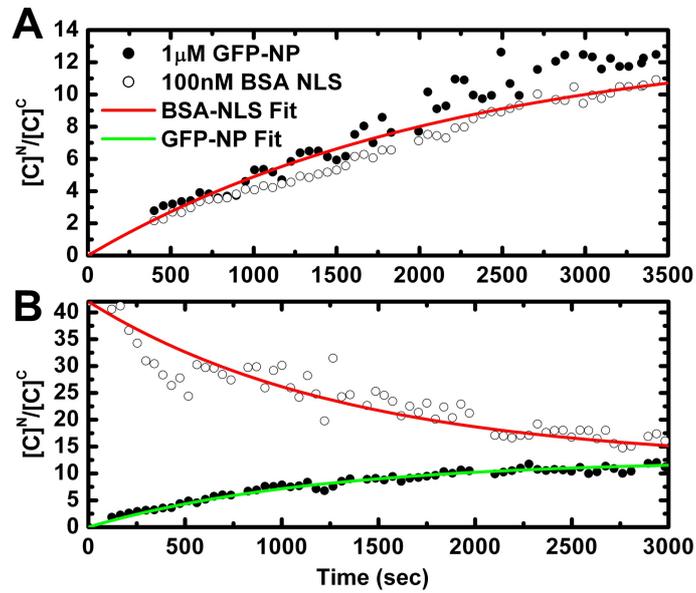

Figure 2

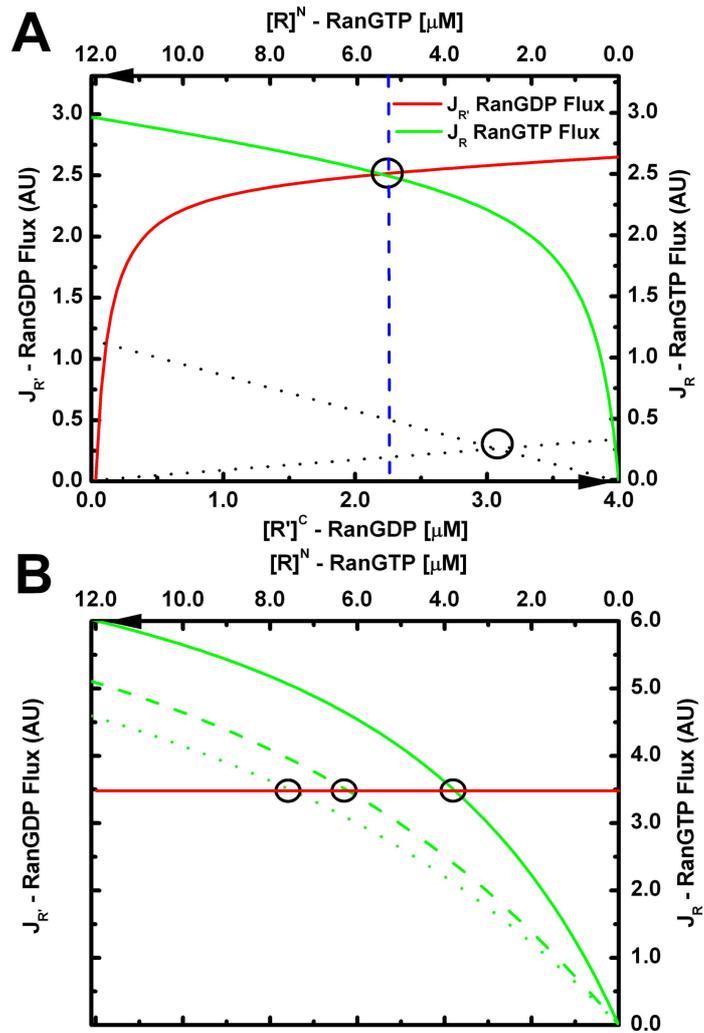

Figure 3

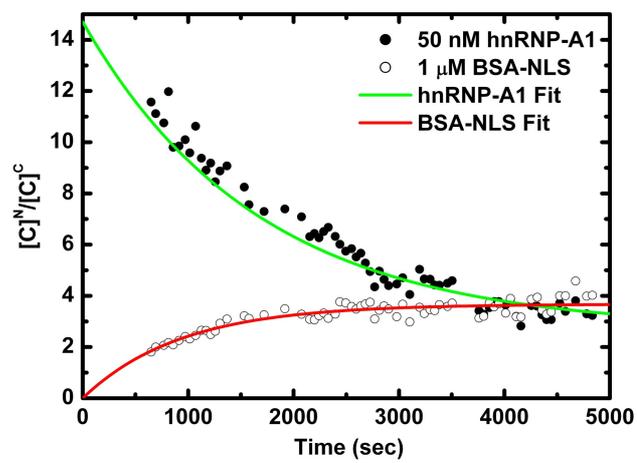

Figure 4

Table 1.

| cytoplasm | pore | nucleus |
|---|---|---|
| $T + C_i \Leftrightarrow TC_i$ | $\longleftrightarrow$ | $TC_i \Leftrightarrow T + C_i$ |
| $T + R' \leftarrow TR$ | $\longleftarrow$ | $TR \Leftrightarrow T + R$ |
| $E + R' \leftarrow ER$ | $\longleftarrow$ | $ER \Leftrightarrow E + R$ |
| $N + R' \Leftrightarrow NR'$ | $\longrightarrow$ | $NR' \rightarrow N + R$ |
| $R'$ | $\longleftrightarrow$ | $R$ |

**Supplementary Information**

We consider the nuclear accumulation kinetics of a new substrate into a transport system that has already reached steady state with regard to the endogenous cargo. Supplementary Figure S1 shows the dependence of initial rate on the affinity of the new probe when added to a fractional concentration of 1% of the total endogenous. The curves are calculated from the equation: $\dfrac{[TC_x]^C}{[TC_e]^C} = \dfrac{[C_x]^C K_e}{[C_e]^C K_x}$. Endogenous affinities may be considered hypothetically uniform, as marked on the graph, or a mean value weighted by the concentration of each endogenous cargo. Time units on the vertical axis are arbitrary as they will depend on the particular permeabilities $p_{TC}$, and the initial rate is just the product $p_{TC_x}[TC_x]^C$. Qualitatively, the new probe will enter the nucleus quickly if it displaces cytoplasmic receptors from the endogenous cargoes, i.e., if its affinity is relatively strong. It will enter slowly if its affinity is weak relative to the endogenous, as in that case it will have little access to the transport receptors in the cytoplasm. The effect is dramatic, and may effectively determine the potential for nuclear accumulation on a relevant time scale. The model prediction bears a striking relation to Fig 5 in (Hodel AE et al, 2006), where nuclear accumulation was measured in yeast as a function of the substrate NLS affinity to receptors.

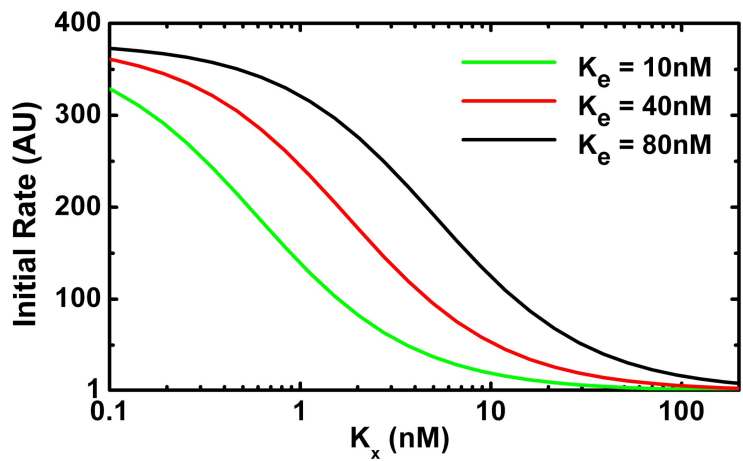

Figure S1